\documentclass[aps,prl,onecolumn, tightenlines, nopacs, amsmath,amssymb,final,letterpaper,12pt]{revtex4}

\usepackage{graphicx}
\usepackage{color}

\usepackage{amssymb,amsfonts,amsmath}
\usepackage{calc}

\begin{document}


\title{Global efficiency of local immunization on complex networks}

\author{Laurent H\'{e}bert-Dufresne}
\author{Antoine Allard}
\author{Jean-Gabriel Young}
\author{Louis J. Dub\'{e}}

\affiliation{D\'epartement de Physique, de G\'enie Physique, et d'Optique, Universit\'e Laval, Qu\'ebec (Qu{\'e}bec), Canada G1V 0A6}


\begin{abstract}
\textbf{Epidemics occur in all shapes and forms: infections propagating in our sparse sexual networks, rumours and diseases spreading through our much denser social interactions, or viruses circulating on the Internet. With the advent of large databases and efficient analysis algorithms, these processes can be better predicted and controlled. In this study, we use different characteristics of network organization to identify the influential spreaders in 17 empirical networks of diverse nature using 2 epidemic models. We find that a judicious choice of local measures, based either on the network's connectivity at a microscopic scale or on its community structure at a mesoscopic scale, compares favorably to global measures, such as betweenness centrality, in terms of efficiency, practicality and robustness. We also develop an analytical framework that highlights a transition in the characteristic scale of different epidemic regimes. This allows to decide which local measure should govern immunization in a given scenario.}
\end{abstract}

\maketitle
Epidemics never occur randomly. Instead, they follow the structured pathways formed by the interactions and connections of the host population \cite{caldarelli,keeling}. The spreading processes relevant to our everyday life take place on networks of all sorts: social (e.g. epidemics \cite{anderson, keeling_05}), technological (e.g. computer viruses \cite{pastor_01a, gomez06} ) or ecological (cascading extinctions in food webs \cite{dunne09}). With a network representation, these completely different processes can be modelled as the propagation of a given agent on a set of nodes (the population) and links (the interactions). Different systems imply networks with different organizations, just as different agents require different epidemic models.

There has long been significant interest in identifying the {\it influential spreaders} in networks. Which nodes should be the target of immunization efforts in order to optimally protect the network against epidemics? Unfortunately, most studies feature two significant shortcomings. Firstly, the proposed methods are often based on optimization or heuristic algorithms requiring nearly perfect information on a static system (e.g. \cite{gallos07,Chen08}); this is rarely the case. Secondly, methods are usually tested on small numbers of real systems using a particular epidemic scenario (e.g. \cite{salathe,masuda}); this limits the scope of possible outcomes. 

\begin{figure}[!b]
 \centering
 \includegraphics[trim = 0mm 0mm 0mm 0mm, clip, width=0.5\linewidth]{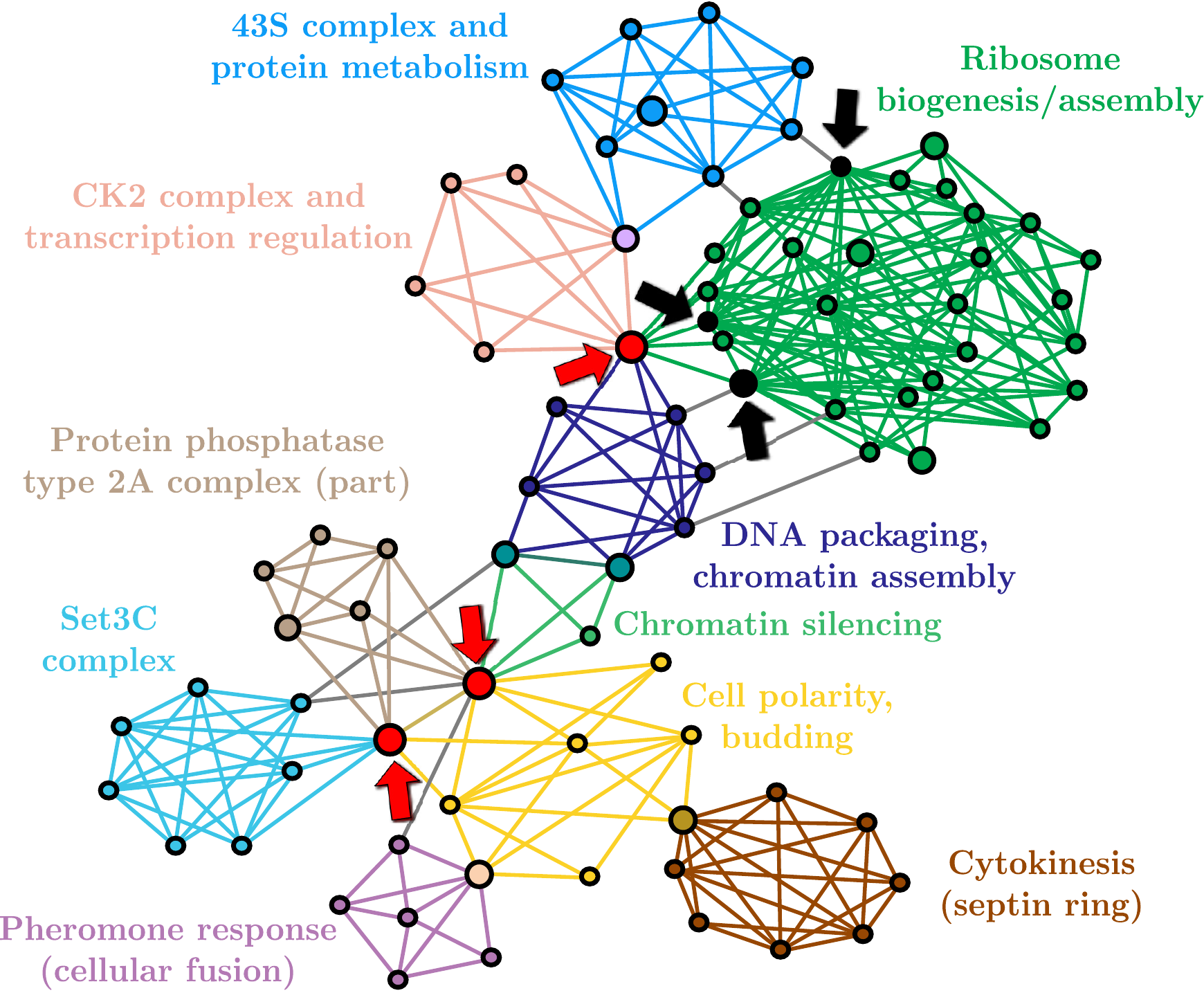}\\
 \caption{\label{Yeast_illustration} \textbf{Protein interactions of {\it S. cerevisiae} (subset)} \cite{palla05}. The three black nodes correspond to the ones with the highest degree, and the three red ones have the highest membership number. In this particular example, it is readily seen that the latter are structurally more influent.}
\end{figure}

We first present a numerical study, perhaps the largest of its kind to date, where we argue that, depending on the nature of the network and of the disease, different immunization tactics have to be taken into consideration. In so doing, we formalize the notion of node influence and illustrate how {\it local knowledge} around a particular node is usually sufficient to estimate its role in an epidemic. We also show how, in certain cases, the influence of a node is not necessarily dictated by its number of connections, but rather by its role in the network's community structure (see Fig.~\ref{Yeast_illustration}). Far from trivial, it follows that an efficient immunization strategy can be obtained solely from local measures, which are easily estimated in practice and robust to noisy or incomplete information. We further develop an analytical formalism ideally suited to test the effects of local immunization on realistic network structures. Combining the insights gathered from the numerical study and this formalism, we finally formulate a readily applicable approach which can easily be implemented in practice.

%
%
%
%
%
\section{Results}
\subsection{Models and measures}
%
There exist two standard models emulating diverse types of epidemics: the {\it susceptible-infectious-recovered} (SIR) and {\it susceptible-infectious-susceptible} (SIS) dynamics. In both, an infectious node has a given probability of eventually infecting each of its susceptible neighbors during its infectious period, which is terminated by either death/immunity leading to the recovered state (SIR) or by returning to a susceptible state (SIS). In the SIR dynamics, for a given transmission probability $T$, the quantity of interest is the mean fraction $R_\textrm{f}$ of recovered nodes once a disease, not subject to a stochastic extinction, has finished spreading (i.e. we focus on the giant component \cite{newman01}). Since each edge can only be followed once, this dynamics investigates how a population is vulnerable to the {\it invasion} of a new pathogen. In the SIS dynamics, we are interested in the prevalence $I^*$ (fraction of infectious nodes) of the disease at equilibrium (equal amounts of infections and recoveries) as a function of the ratio $\lambda=\alpha/ \beta$ of infection rate $\alpha$ and recovery rate $\beta$. This particular dynamics permits the study of how a given network structure can {\it sustain} an already established epidemic.

Should a fraction $\varepsilon$ of the population be fully immunized, our objective is to identify the nodes whose absence would minimize $R_\textrm{f}$ and $I^*$. The {\it epidemic influence} of a node --- that is the effect of its removal on $R_\textrm{f}$ and $I^*$ --- depends mainly on its role in the organization of the network. Hence to efficiently immunize a population, we must first understand its underlying structure.

Network organization can be characterized on different scales, each of which affect the dynamics of propagation. At the microscopic level, the most significant feature is the {\it degree} of a node (its number of links, noted $k$) which in turn defines the degree distribution of the network. The significance of the high-degree nodes (the {\it hubs}) for network structure in general \cite{barabasi99}, for network robustness to random failure \cite{barabasi00b} and for epidemic control \cite{pastorsatorras02} has long been recognized.

At the macroscopic level, the role of a node can be described by its {\it centrality}, which may be defined in various ways. Frequently used in the social sciences is the {\it betweenness centrality} ($b$), quantifying the contributions of a given node to the shortest paths between every pair of nodes in the network \cite{freeman79}. Arguably, this method should be among the best estimate of a node's epidemic influence as it directly measures its role in the different pathways between all other individuals \cite{barthelemy04}, yet at a considerable computational cost. A simpler method, the k-core (or k-shell) decomposition \cite{batagelj02, batagelj03}, assigns nodes to different layers (or {\it coreness} $c$) effectively defining the core and periphery of a network (high and low $c$ respectively). It has recently been shown that coreness is well suited to identify nodes that are the most at risk of being infected during the course of an epidemic \cite{kitsak10}. In light of our results, we will be able to discuss the distinction between a node's vulnerability to infection and its influence on the outcome of an epidemic.

The mesoscopic scale has recently been the subject of considerable attention. At this level of organization, the focus is on the redundancy of connections forming dense clusters referred to as the community structure of the network \cite{ahn, palla05}. Nodes can be distinguished by their {\it membership} number $m$, i.e., the number of communities to which they belong. We will consider that two links of a given node are part of one community if the neighbours they reach lead to significantly overlapping neighbourhoods \cite{ahn}. This definition is directly relevant to epidemic dynamics as links within communities do \textit{not} lead to new potential infections. We call {\it structural hubs} the nodes connecting the largest number of different communities. These nodes act as bridges facilitating the propagation of the disease from one dense cluster to another. Targeting structural hubs to hinder propagation in structured populations has been previously proposed and investigated \cite{salathe,masuda}, but has yet to be tested extensively.

Note that the microscopic and mesoscopic levels (as defined above) are characterized by {\it local measures} in the sense that they do not require a complete knowledge of the network, in contrast to {\it global measures} like the betweenness centrality. Moreover, as we will see, \emph{local measures are less sensitive to incomplete or incorrect information}. Adding, removing or rewiring a link only affects the degree or membership of nodes directly in the neighbourhood of the modification; whereas the same alterations can potentially affect the centrality of nodes anywhere in the network through cascading effects. Furthermore, even if community detection often requires the tuning of a global resolution parameter, we will see that this additional step does not affect the identification of structural hubs, meaning that local information is sufficient to accurately determine a node's memberships.

In our numerical simulations we will have a perfect knowledge of static networks. This will allow us to use global measures as a reference to test the efficiency of local measures best suited in practice. We therefore ask without discrimination: which of the degree $k$, the coreness $c$, the betweenness centrality $b$ or the membership number $m$ is the best identifier of the most influential nodes on the outcome of an epidemic? To answer this question, we have simulated SIR and SIS dynamics with Monte Carlo calculations on 17 real-world networks. In each case, a fraction $\varepsilon$ of the nodes was removed in decreasing order of the nodes' score for each of the four different measures. By comparing their efficiency to reduce $R_\textrm{f}$ or $I^*$ as a function of $\varepsilon$, we are able to establish which measure is best suited for a given scenario characterized by a network structure, a propagation dynamics and a disease transmissibility (i.e. probability of transmission).
%
%
%
%
%
\subsection{Case study: a data exchange network}
%
We first illustrate our methods using the network of users of the Pretty-Good-Privacy algorithm for secure information interchange (hereafter, the PGP network) \cite{boguna04}, which could be the host of the propagation of computer viruses, rumors or viral marketing campaigns. Results for the 16 other networks are presented and discussed in the next section as well as in the Supporting Information (SI) document.

\begin{figure}[!b]
\centering
\includegraphics[width=0.5\linewidth]{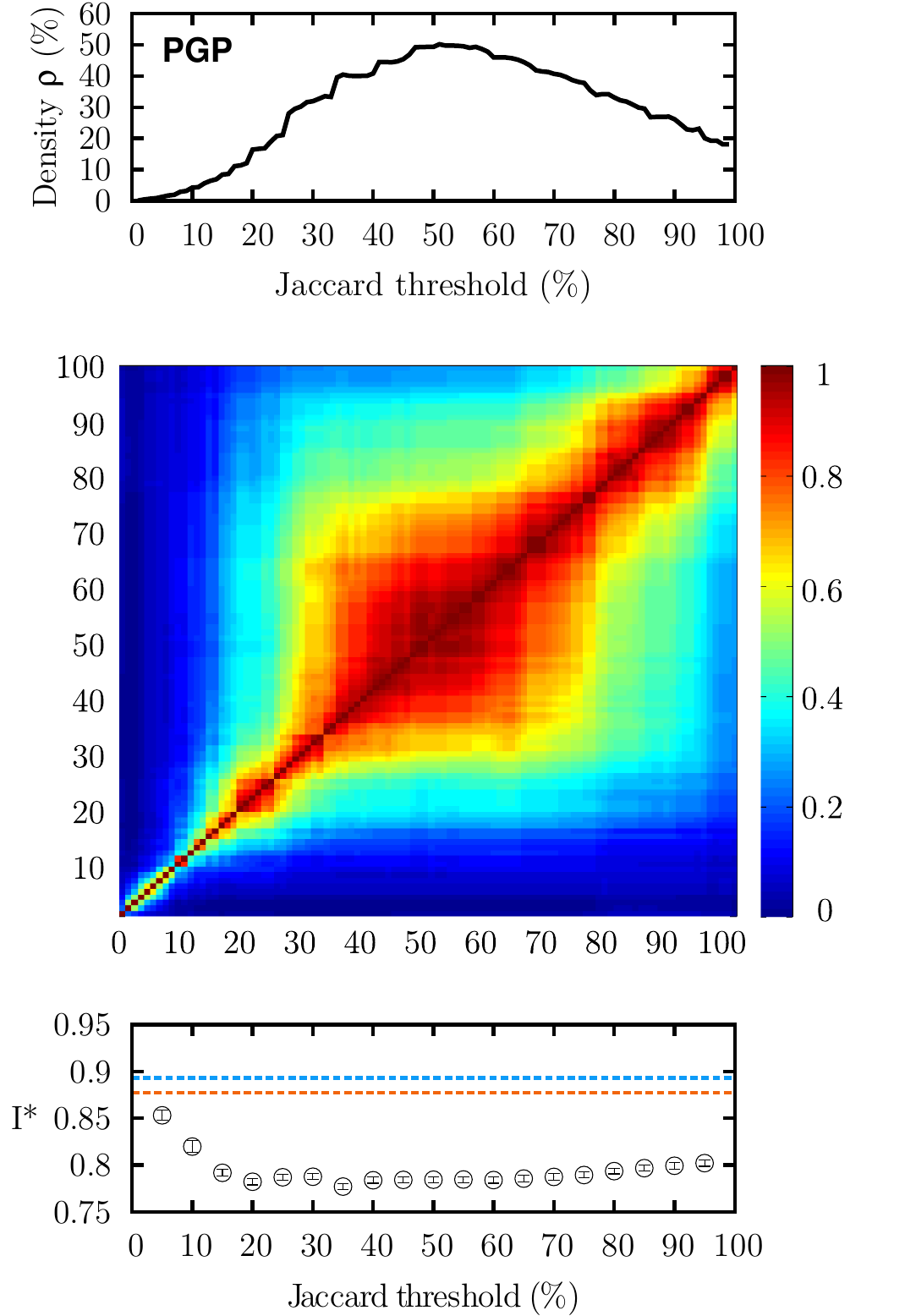}
\caption{\label{PGP_comm} \textbf{Robustness of structural hubs in the PGP network.} (top) Community density ($\rho$) obtained through different Jaccard thresholds. (middle) Robustness of the structural hubs identification methods. Element $(i,j)$ gives the overlap (normalized) between the structural hubs (top $1\%$) selected with thresholds $i$ and $j$. The highest line and last column of the matrix correspond to the case where the membership number equals the degree. (bottom) Prevalence $I^*$ of SIS epidemics with $\lambda = 5$ when the top $1\%$ of structural hubs are removed (compared with the results without removal in blue or with random targets in orange).}
\end{figure}

Communities in the network are extracted with the link community algorithm of Ahn {\it et al.} \cite{ahn}. This algorithm groups links --- and therefore the nodes they join --- into communities based on the overlap of their respective neighbouring nodes. It is this overlap that reduces the number of new potential infections in a community structure, as opposed to a random network. This method thus reflects our understanding of how communities affect disease propagation. While it may not directly detect the social groups or functional modules of a network, it identifies significant clusters of redundant links. This redundancy or overlap is quantified through a Jaccard coefficient, and two links are grouped into the same community when their coefficient exceeds a certain threshold. The threshold value acts as a resolution, enabling to look at different levels of organization. As suggested in \cite{ahn}, the value of the threshold is chosen to maximize the average density $\rho$ of the communities (see Material and Methods). As this choice may seem arbitrary, Fig.~\ref{PGP_comm} investigates the similarity between the nodes with the highest membership numbers, for different thresholds. It suggests that the membership number is fairly robust around the threshold. Moreover, Fig.~\ref{PGP_comm} also demonstrates that the effect of the removal of the structural hubs on a SIS epidemics is very robust to the choice of the threshold. Thus, we will henceforth use the membership numbers obtained with the threshold value corresponding to the highest community density.

\begin{figure}[!b]
\centering
\includegraphics[trim = 0mm 0mm 0mm 0mm, clip, width=0.5\linewidth]{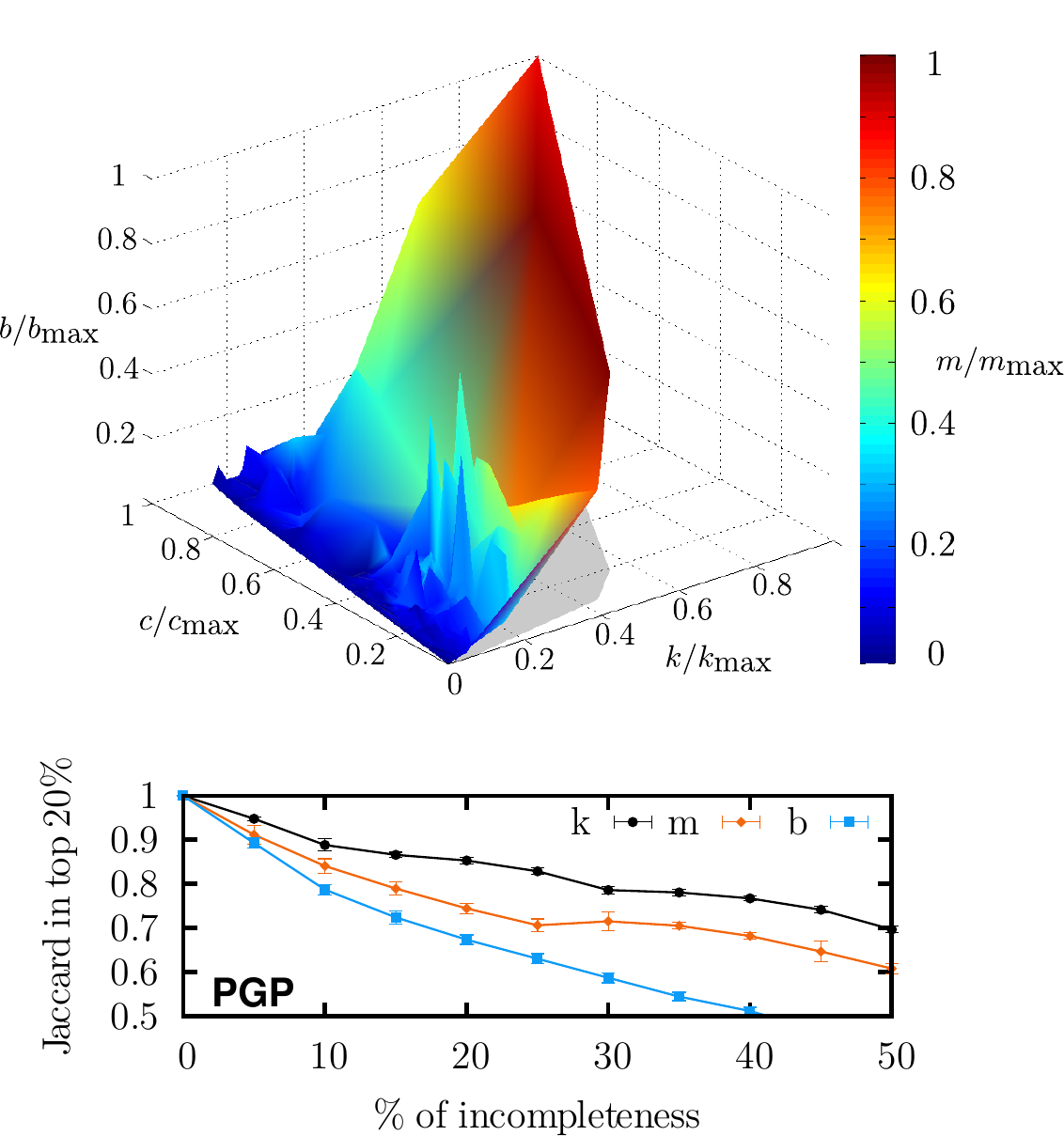}\\
\caption{\label{PGP_cloud} \textbf{Difference in immunization targets for the PGP network.} (top) We present correlations between the degree ($k$, right axis), the coreness ($c$, left axis), the betweenness centrality ($b$, vertical axis) and the membership number ($m$, color) for each nodes. Each measure is normalized according to the highest value found in the network. Each node is represented in this 4-dimensional space and a simple triangulation procedure then yields a more intelligible appearance. Structural hubs (dark red) can be found even at relatively small degree ($\sim k_{\textrm{max}}/2$), coreness ($\sim c_{\textrm{max}}/5$) and centrality ($\sim b_{\textrm{max}}/3$). (bottom) Jaccard coefficient between the ensemble of nodes identified as part of the top $20\%$ according to a given measure ($k$, $m$ or $b$) on two versions of the network: the original complete network and a network ensemble where a certain percentage of links has been randomly removed (horizontal axis). The shorter the range of a measure, the more robust it is to incomplete information.}
\end{figure}

The differences, if any, between the efficiency of the different methods are due to the immunized nodes not being the same. Figure \ref{PGP_cloud}(top) investigates the correlations between the different properties ($k$, $b$, $c$ and $m$) of each node. Perhaps the most important result here is that nodes with a high membership number may have relatively small degree, coreness and betweenness centrality. Hence, we expect the immunizing method based on community structure to have a different influence on the outcome of epidemics. Figure \ref{PGP_cloud}(bottom) shows the consistensy (or lack thereof) of a given measure, depending on the quality of the available data. The robustness of local (micro and meso) measures is of obvious practical advantage. Both robustness and correlations are further investigated in the SI.

To study various epidemic scenarios, we consider both SIS and SIR dynamics (which may behave quite differently) with different values of the transmission probability ($\lambda$ and $T$ for SIS and SIR, respectively). In fact, each network features an {\it epidemic threshold}, i.e. critical values $\lambda _c$ \cite{hebert10} and $T_c$ \cite{newman02}, below which $I^*$ and $R_\textrm{f}$ vanish to zero in an equivalent infinite network ensemble. As we will show, the observed behavior can differ significantly depending whether or not $\lambda$ and $T$ are close to their critical value.

\begin{figure}[!b]
\centering
\includegraphics[width=0.5\linewidth]{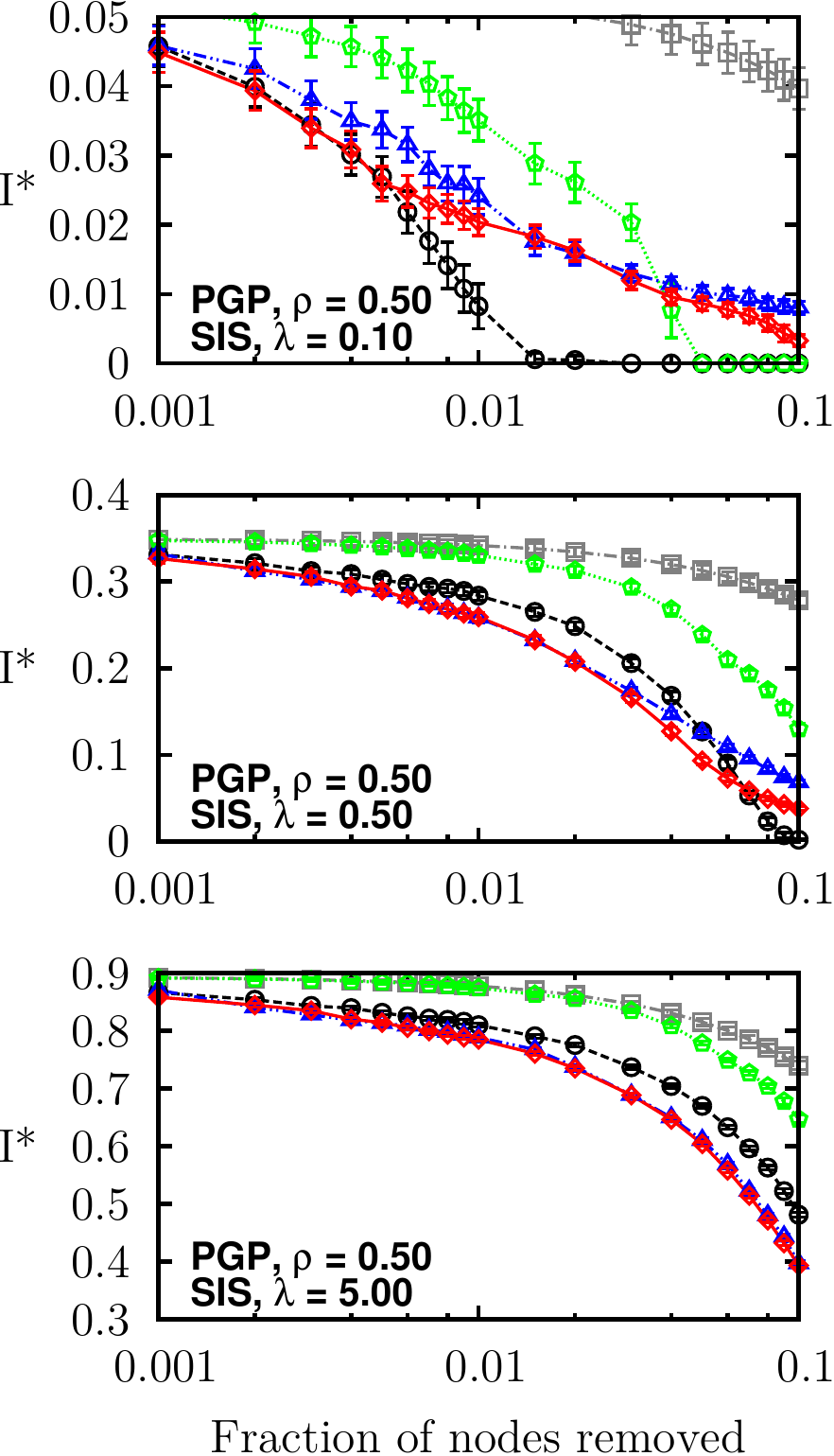}
\caption{\label{PGP_SIS} \textbf{Efficiency of the immunization methods against an SIS epidemics on the PGP network}. Nodes are removed in decreasing order of their score according to each method: coreness (green pentagons), degree (black circles), betweenness centrality (blue triangles) and memberships (red diamonds) and the effect of removal is then quantified in terms of the decrease of the prevalence $I^*$. The prevalence of the epidemics when the removed nodes are chosen at random (grey squares) has been added for comparison. Figures are presented in increasing order of transmissibility ($\lambda$) from top to bottom.}
\end{figure}

Figure \ref{PGP_SIS} presents results of different immunization methods against SIS dynamics for different values of $\lambda$. On the top figure, where $\lambda$ is near $\lambda _c$, the most successful method of intervention is to target nodes according to their degree. At low transmissibility, the disease follows only a very small fraction of all links. The shortest paths are seldom used and the poor performance of betweenness centrality follows. Moreover, the disease will not be affected by the community structure, because even in dense neighbourhoods, most links will not be travelled. We then say that the disease, unaffected by link clustering, follows a tree-like structure (without loops), where community memberships are insignificant. It is therefore better to simply remove as many links as possible.

As $\lambda$ increases beyond $\lambda_c$, we see that immunization based on membership numbers quickly outperforms the other methods. As more links are travelled, the disease is more likely to follow superfluous links in already infected communities. Hubs sharing their many links within few communities are therefore not as efficient in causing secondary infections as one might expect. Similarly, targeting through betweenness centrality also performs better with higher $\lambda$, albeit not as well as membership-targeting in this case. For $\lambda \gg \lambda _c$, immunization based on membership numbers (local) and on betweenness centrality (global) converge toward similar efficiency, significantly outperforming degree-based immunization.

\begin{figure}[!b]
\centering
\includegraphics[trim = 0mm 0mm 0mm 0mm, clip, width=0.4\linewidth]{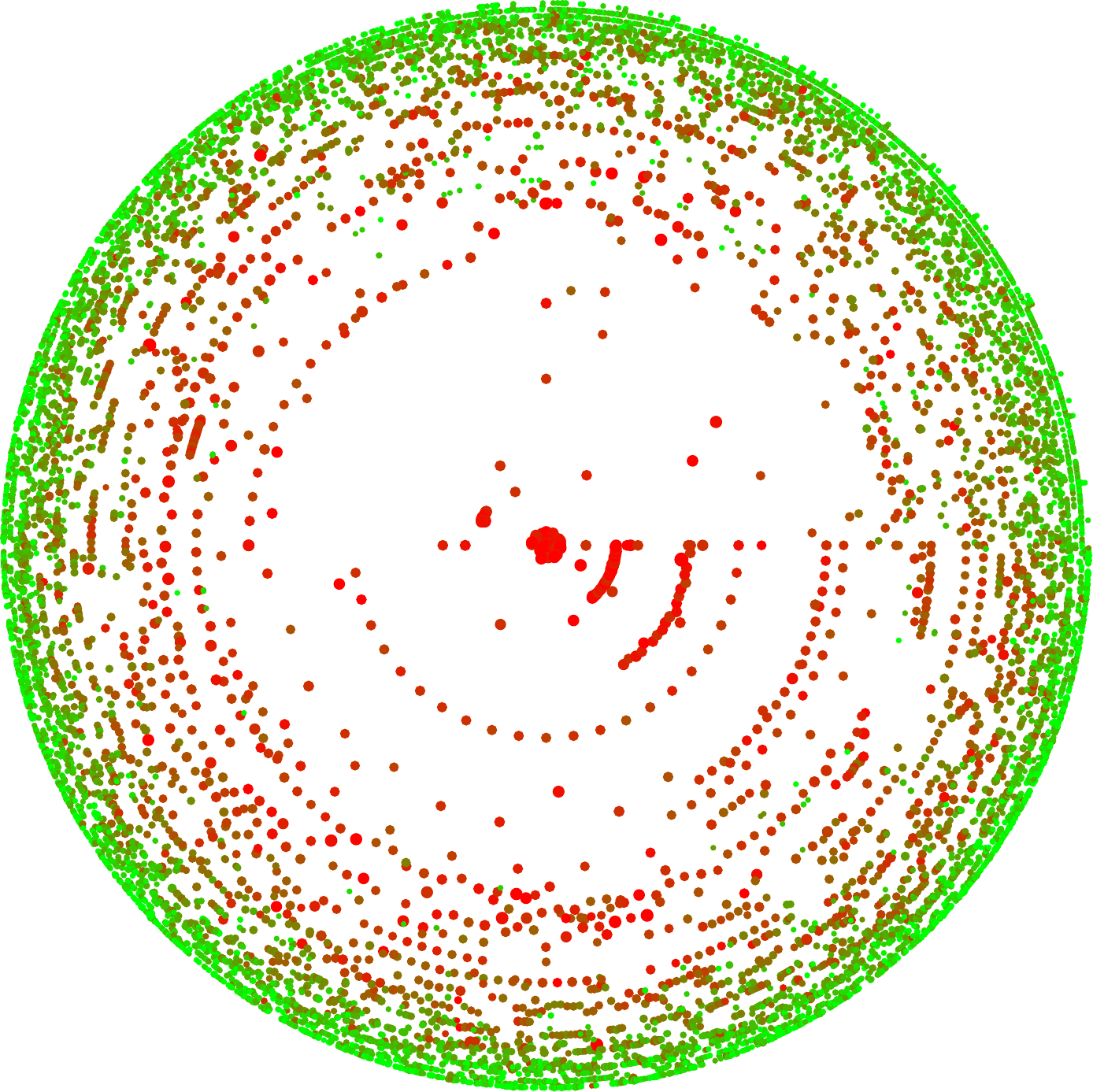}\\
\caption{\label{PGP_cores} \textbf{k-core decomposition of the PGP network.} Representation (based on \cite{alvarez06}) of the k-shells in the PGP network with nodes colored according to their total infectious period during a given time interval. Red nodes are more likely to be infectious at any given time than green nodes as the color is given by the square of the fraction of time spent in infectious state. Note how the central nodes (the core) of the network are most at risk.}
\end{figure}

Another interesting feature of our results is the poor performance of immunization based on node coreness. A previous study had clearly shown that epidemics mostly flourished within the core of the network (see Fig. \ref{PGP_cores}) because of its density \cite{kitsak10}. Ironically, this density also implies redundancy. While the core nodes are highly at risk of being infected, their removal has a limited effect because there exist alternative paths within their neighbourhood: the core offers a perfect environment to the disease and is consequently robust to node removal. It is therefore more effective to stop the disease from reaching, or leaving, the core by removing the nodes bridging other neighbourhoods (i.e. the structural hubs).

Similar conclusions are drawn for the SIR dynamics. As $T$ moves away from $T_c$, the most significant level of organisation shifts from the degree (microscopic) to communities (mesoscopic) as membership-based immunization progressively outperforms the other strategies.

\begin{figure*}[p!]
\centering
\includegraphics[trim = 0mm 0mm 0mm 0mm, clip, width=0.76\linewidth]{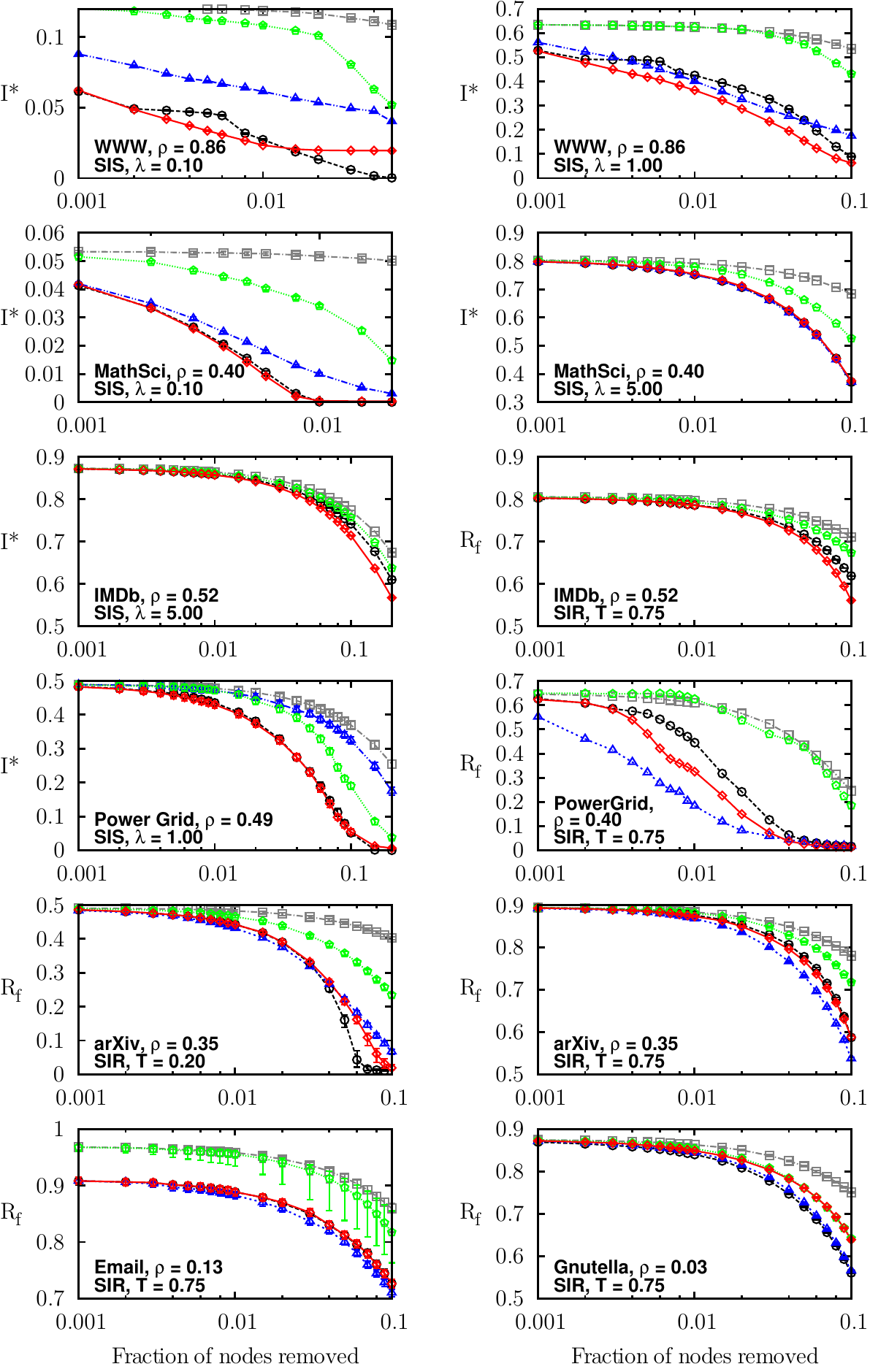}
\caption{\label{All_results} \textbf{Efficiency of the immunization methods against SIS and SIR epidemics on several networks.} Nodes are removed in decreasing order of their score according to each method: coreness (green pentagons), degree (black circles), betweenness centrality (blue triangles) and memberships (red diamonds) to measure efficiency by the decrease of $I^*$ or $R_\textrm{f}$. The size of the epidemics for random removal of nodes (gray squares) is added for comparison. Error bars have been omitted for clarity of the SIR results on the Power Grid, but are shown in the SI.}
\end{figure*}

%
%
%
%
%
%
\subsection{Results on networks of diverse nature}
%
In this section, we highlight different behaviours observed in social, technological and communication networks using 7 other datasets (full results for the 17 datasets are available in the SI): subset of the World Wide Web (WWW) \cite{barabasi99}, MathSciNet co-authorship network (MathSci) \cite{palla_08}, Western States Power Grid of the United States (Power Grid) \cite{strogatz}, Internet Movie Database since 2000 (IMDb) \cite{lhd_prl}, cond-mat arXiv co-authorship network (arXiv) \cite{palla05}, e-mail interchanges between members of the University Rovira i Virgili (Email) \cite{guimera03} and Gnutella peer-to-peer network (Gnutella) \cite{leskovec}.

The results for the WWW, MathSci and IMDb networks further support our previous conclusions, with the exception that membership-based immunization performs surprisingly better than the degree-based variant even near the epidemic threshold of the network (see WWW and MathSci). The betweenness-centrality-based immunization was not tested on IMDb because of computational constraints (its computation required over 800 hours with our available ressources and a standard algorithm \cite{brandes}), which illustrates a significant limit of this measure. Approximations could have been used \cite{madduri}, but the intricate (and mostly unknown) relationship between the efficiency of the measure and the accuracy of the approximation would have only caused additional uncertainties.

The results presented for the Power Grid network illustrate a fundamental difference between the SIS and the SIR dynamics: while we are interested in the fraction of the network sustaining an established epidemic in SIS, it is the fraction of nodes invaded by a new disease that is relevant in SIR. In fact, the structure of the Power Grid, a chain of small, easily disconnected modules, enhances the qualitative discrepancy between the epidemic influence of nodes subjected to these two dynamics. For the SIS dynamics, the membership-based intervention is the most efficient because it weakens all modules, limiting the prevalence of the disease. In distinction, targeting through betweenness centrality merely separates the modules, so that they indiviually remain infected. For the SIR dynamics, separating the modules is the best approach as it directly stops the infection from spreading; while weakened -- but connected -- modules still provide pathways. This effect is a direct consequence of the particular structure of the Power Grid and is insignificant on other networks.

Finally, the last set of results, on arXiv, Email and Gnutella, present the effect of the community density $\rho$ on the performance of membership-based immunization. For very small $\rho$, the paths within communities do not qualitatively differ from the links bridging neighborhoods in their effect on the disease propagation. This targeting method is therefore expected to converge toward degree-based immunization if $m$ and $k$ are strongly correlated. However, as most tested networks had fairly dense communities, $\rho \geq 0.3$, the relevance of memberships should not be understated.
%
%
%
%
%
%
\subsection{Investigation of the epidemic regimes transition}
%

The results of the previous sections suggest that local information (i.e., degree, membership) is often sufficient for a nearly optimal global immunization. More precisely, we found these two methods to outperform or to be as efficient as the betweenness centrality (the global method used for comparison) in 62 of the 68 studied scenarios (i.e., 17 networks / 2 dynamics / 2 transmissibility regimes). This implies that membership (e.g., on PGP), degree (e.g., Gnutella) or both (e.g. MathSci) lead to an immunization at least as efficient as global methods while having the noteworthy advantage of requiring much less information and of being less sensitive to incomplete information. This section focuses on the conditions guiding the choice between the degree-based or the membership-based immunization strategy. In this respect, Figs. 4 and 6 provide a useful hindsight: the membership-based strategy is more efficient than the degree-based one when transmissibility is high and/or when communities are dense. To further our understanding and test this hypothesis, we introduce a random network model featuring a community structure, and exactly solve its final state ($R_\mathrm{f}$) under SIR dynamics using generating functions.

\begin{figure}[!b]
 \centering
 \includegraphics[trim = 0mm 0mm 0mm 0mm, clip, width=0.5\linewidth]{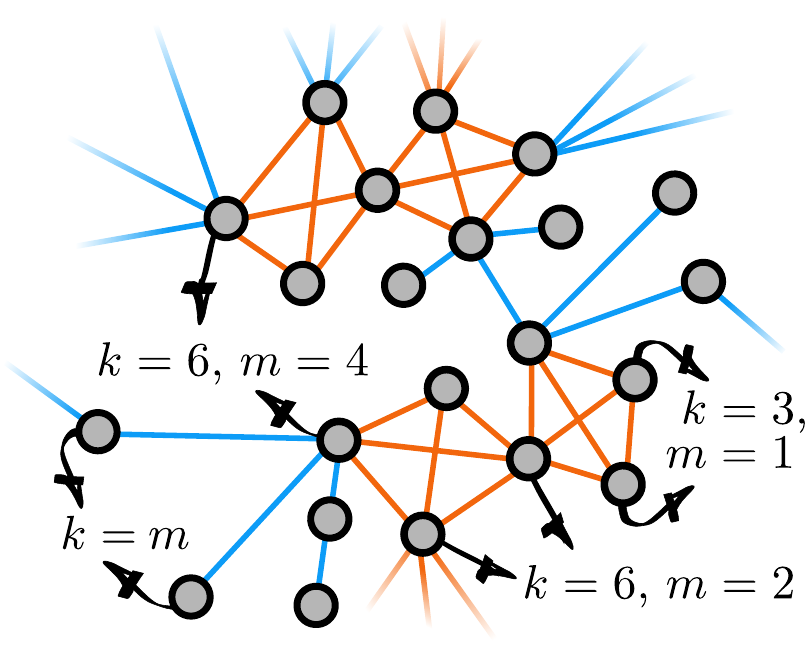}
 \caption{\label{InSilico} \textbf{Synthetic networks with tunable community structure.} Orange links belong to motifs of size $M=4$, and single links are shown in blue. The degree $k$ and membership $m$ of a few selected nodes are indicated. They belong to $i = (k-m)/(M-2)$ motifs and have $j = \left[(M-1)m-k\right]/(M-2)$ single links.}
\end{figure}

Our model is a slightly modified version of the configuration model \cite{newman01,newman03} where nodes are connected either through single links or through motifs (see Fig.~\ref{InSilico} for an example). Motifs are used to simulate the effect of a community structure, that is the redundancy of the neighbourhoods of nodes. Our motifs are composed of $M$ nodes, all connected to each other, and a node belongs to $i$ motifs and has $j$ single links with probability $p(i,j)$. This node therefore has a degree ($k$) equal to $(M\!-\!1)i\!+\!j$ and a membership ($m$) equal to $i\!+\!j$. Networks are generated using a stub pairing scheme: a node belonging to $i$ motifs and having $j$ single links has $i$ ``motif stubs'' and $j$ ``link stubs''. Groups and single links are then formed by randomly choosing $M$ motif stubs and 2 link stubs, respectively, and then by linking the corresponding nodes to one another. This last step is repeated until none of the motif and link stubs remains. The distribution $\{p(i,j)\}_{i,j \in \mathbb{N}}$ therefore defines a maximally random network ensemble, and the results obtained are averaged over this ensemble.

\begin{figure}[!b]
 \centering
 \includegraphics[trim = 0mm 0mm 0mm 0mm, clip, width=0.5\linewidth]{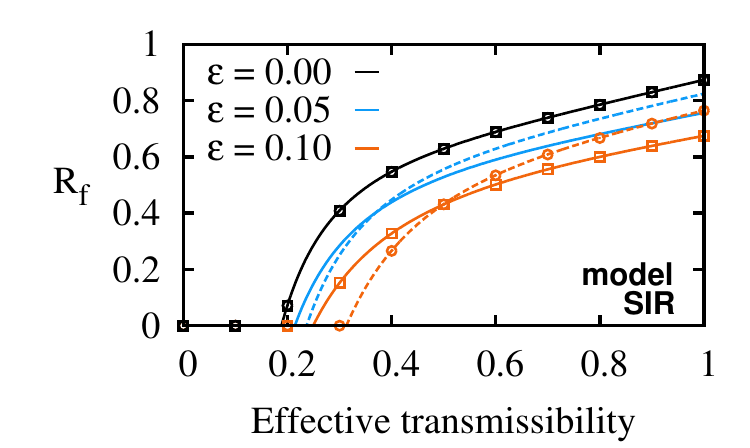}
 \caption{\label{Anall} \textbf{Results of local immunization methods on synthetic networks.} Final sizes of SIR epidemics after immunization of various fractions $\varepsilon$ of nodes on synthetic networks with $M=4$ and an heterogeneous degree distribution (details in SI). Near the epidemic threshold, targeting by degree (dotted curves) is the better choice whereas targeting by memberships (solid curve) should be preferred for higher transmissibility. Monte Carlo simulations were also performed to validate the formalism and indicated on the curves (the case $\varepsilon = 0.05$ is omitted not to clutter the graph) with circles (targeting by degree) and squares (targeting by membership).}
\end{figure}

Extending previous work \cite{allard12b}, we compute the expected value of $R_\textrm{f}$ for the network ensemble just defined where nodes and links are randomly removed to simulate immunization and disease transmission (SIR dynamics), respectively. Full details are given in the SI. Using typical values for $\{p(i,j)\}$, our model illustrates and confirms our hypothesis by clearly showing in Fig.~\ref{Anall} a transition of efficiency between the degree-based and the membership-based immunization strategy. Initially less efficient when the transmissibility is low
(i.e., higher threshold, lower value of $R_\textrm{f}$), membership progressively outperforms degree as the transmissibility increases. As mentionned above, for lower values of $T$, the best option is therefore to immunize the hubs (high $k$) to shift the degree distribution towards lower degrees. For higher values of $T$, targeting structural hubs (high $m$) that act as bridges between ``independent'' neighbourhoods leads to a more efficient immunization as it reduces the number of paths between different regions of the network. Note that we do not explicitly model the effect of community density. This could have been done by letting links exist independently with a given probability
$\eta$. This is however identical to letting the disease propagate with probability $\eta T$. Thus, the value of $T$ in Fig.~\ref{Anall} is related to the density of the communities, and our conclusions can therefore be extended the cases of low/high community densities.

%
%
%
%
%
\section{Discussions}
%

One of the main contributions of this work is to offer a formal definition of the epidemic influence of nodes, i.e. the effect of its removal on $I^*$ of $R_\textrm{f}$, which is open to diverse methods of approximation. Our results confirm that standard measures such as the degree or betweenness centrality are \emph{not always} the best indicators of a node's influence. Moreover, we have highlighted that the coreness, which has recently been proposed as an indicator of nodes' influence \cite{kitsak10}, offers poor performances. This has brought us to distinguish between individual risk and global influence. We have also illustrated how a universal approach is still wanting, since different networks and different diseases require different methods of intervention.

Consequently, the fact that the numbers of links and/or communities to which a node belongs are excellent measure of its epidemic influence --- at times better, at times equivalent, but never much worse than global centrality measures --- is a particularly important result. The fact that they both are \emph{local measures} is especially relevant considering that we rarely have access to the exact network structure of a system, either because it is simply too large (WWW), too dynamic (email networks) or because the links themselves are ill-defined (social networks). Not only are local measures computable from a limited subset of a network (which is often the only available information), but a coarse-grained measure like membership is even more interesting as it is easier to estimate than a node's actual degree. For instance, consider how much simpler it is to enumerate your social groups (work, family, etc.) than the totality of your acquaintances.

Finally, the existence of a transition between two epidemic regimes with different characteristic scales may well be the single most important conclusion of this work. In the first regime, for low transmissibility and sparse communities, the microscopic structural features (i.e. node connectivity or degree) offer the most relevant information; while for higher transmissibility and denser communities, mesoscopic features (i.e node communities or membership) appear more relevant. We expect to see an equivalent transition between any pair of measures which oppose the micro and meso scales (e.g. different range-limited measures of centrality \cite{Ercsey-Ravasz}).

Based on our empirical and analytical results, we thus propose a simple procedure on how to judge which local measure can be expected to yield the best results in a given situation. From the available subset of a given network:
\begin{enumerate}
  \item Obtain the degree distribution to estimate the transmissibility of the disease in relation to the epidemic threshold $\lambda _c$ \cite{hebert10} or $T_c$ \cite{newman02}.
  \item If easily transmissible ($\lambda \gg \lambda_c$ or $T \gg T_c$), evaluate the network's community structure; otherwise, go to 4.
  \item If the community density is high ($\rho \gtrsim 0.3$), immunize nodes according to their memberships; otherwise, go to 4.
  \item For a transmissibility near the epidemic threshold, or for sparse communities (low $\rho$), immunize according to the degree of the nodes.
\end{enumerate}

The analytical and numerical frameworks used in this work are expected to guide immunization efforts toward simpler, more precise and efficient strategies. Likewise, the introduction of a node influence classification scheme opens a new avenue for finding better local estimates of a node's role in the global state of its system.

\section{materials}
\small{\paragraph*{\textbf{Betweenness centrality}} For all pairs $(a,b)$ of nodes excluding $i$, list the $n_{a,b}$ shortest paths between $a$ and $b$. Let $n_{a,b}(i)$ be the number of these paths containing $i$. The betweenness centrality $b_i$ of node $i$ is then given by:
\begin{equation}
b_i = \sum _{(a,b)} \frac{n_{a,b}(i)}{n_{a,b}} \; .
\end{equation}

\paragraph*{\textbf{Coreness}} The coreness of node $i$ is the highest integer $c_i$ such that the node is part of the set of all nodes with at least $c_i$ links within the set.

\paragraph*{\textbf{Community detection}} Two links, $e_{ij}$ and $e_{ik}$, from a given node $i$, are said to belong to the same community if their Jaccard coefficient $J(e_{ij},e_{ik})$ (similarity measure) is above a given threshold $J_c$ :
\begin{equation}
J\left(e_{ij},e_{ik}\right) = \frac{n_+(j) \cap n_+(i)}{n_+(j) \cup n_+(i)} > J_c \ ,
\end{equation}
where $n_+(u)$ is the set containing the neighbors of $u$ including $u$.

\paragraph*{\textbf{Community density}} The density $\rho _i$ of a community $i$ of $n_i>2$ nodes and $d_i$ links is the proportion of the possible redundant links that do exist; i.e., the fraction of existing links excluding the minimal $n_i-1$ links that are needed for this community to be connected:
\begin{equation}
\rho _i = \frac{d_i - (n_i-1)}{\frac{n_i(n_i-1)}{2} - (n_i-1)} \ .
\end{equation}
The community density $\rho$ is then calculated according to
\begin{equation}
 \rho = \frac{1}{D} \sum_i d_i \rho_i \ ,
\end{equation}
where $D$ is the total number of links not belonging to single link communities, for which $\rho_i =0$ \cite{ahn}.

\paragraph*{\textbf{Immunization}} To perform the immunization of a fraction $\varepsilon$ of the network according to a certain measure $\Gamma$, we remove the $\varepsilon N$ nodes with the highest $\Gamma$. When a choice must be made (nodes with equal $\Gamma$), all decisions are taken randomly and individually for each simulated epidemics.

\paragraph*{\textbf{Monte Carlo simulations}} To investigate the fraction of a network which can {\it sustain} an epidemics, SIS simulations start with all nodes in an infectious state and are then relaxed until an equilibrium is reached. To investigate the mean fraction of a network which a disease can {\it invade}, SIR simulations start with a single randomly chosen infectious node and run until there are no more infectious nodes. Results shown in the figures are obtained by averaging over the outcome of several numerical simulations until the minimal possible standard deviation (limited by network structure and finite size) is obtained. For the SIR dynamics, only the simulations leading to large-scale epidemics (at least 1\% of the nodes) were considered. The complete procedure is given in the SI.
%




\end{document}